\documentclass[12pt,prd,tightenlines,nofootinbib,showpacs]{revtex4}
\newcommand{\be}{\begin{equation}}
\newcommand{\ee}{\end{equation}}
\newcommand{\bdis}{\begin{displaymath}}
\newcommand{\edis}{\end{displaymath}}
\newcommand{\bga}{\begin{equation}\begin{gathered}}
\newcommand{\ega}{\end{gathered}\end{equation}}
\usepackage{bm}
\usepackage{graphics}
\usepackage{rotating}
\usepackage{epsfig}
\usepackage{amsmath}
\usepackage{amsfonts}

\begin{document}
\title{\begin{flushright}{\rm\normalsize SSU-HEP-14/02\\[3mm]}\end{flushright}
Radiative nonrecoil nuclear finite size corrections of order
$\alpha(Z\alpha)^5$ to the hyperfine splitting of $S$-states in muonic hydrogen}
\author{\firstname{R.~N.} \surname{Faustov}}
\affiliation{Dorodnicyn Computing Centre, Russian Academy of Science, Vavilov Str. 40, 119991, Moscow, Russia}
\author{\firstname{A.~P.} \surname{Martynenko}}
\affiliation{Samara State University, Pavlov Str. 1, 443011, Samara, Russia}
\affiliation{Samara State Aerospace University named after S.P. Korolyov, Moskovskoye Shosse 34, 443086,
Samara, Russia}
\author{\firstname{G.~A.} \surname{Martynenko}}
\author{\firstname{V.~V.} \surname{Sorokin}}
\affiliation{Samara State University, Pavlov Str. 1, 443011, Samara, Russia }

\begin{abstract}
On the basis of quasipotential method in quantum electrodynamics
we calculate nuclear finite size radiative corrections of order
$\alpha(Z\alpha)^5$ to the hyperfine structure of $S$-wave energy levels in
muonic hydrogen and muonic deuterium. For the construction of the particle
interaction operator we employ the projection operators on the particle bound states
with definite spins. The calculation is performed in the infrared safe Fried-Yennie
gauge. Modern experimental data on the electromagnetic form factors of
the proton and deuteron are used.
\end{abstract}


%\keywords{Hyperfine structure, muonic atoms, quantum electrodynamics.}

\maketitle

In last years a significant theoretical interest in the investigation of fine and hyperfine
energy structure of simple atoms is related with light muonic atoms: muonic hydrogen,
muonic deuterium and ions of muonic helium. This is generated by essential progress achieved
by experimental collaboration CREMA (Charge Radius Experiment with Muonic Atoms)
in studies of such simple atoms \cite{CREMA}. The measurement of the transition frequency
$2S^{f=1}_{1/2}-2P^{f=2}_{3/2}$ in muonic hydrogen
leads to a new more precise value of the proton charge radius.
For the first time the hyperfine splitting of $2S$ state in muonic hydrogen was measured.
Analogous measurements in muonic deuterium are also carried out and planned for the publication.
It is important to point out that the CREMA experiments set a task to improve by an order of
the magnitude numerical values of charge radii of simplest nuclei (proton, deuteron,
helion and $\alpha$-particle). Successful realization of such program is based on precise theoretical
calculations of different corrections to the energy intervals of fine and hyperfine structure
of muonic atoms \cite{egs,borie1,kp1996,ibk}. Nuclear structure corrections play in this investigation
a special role and, possibly, can solve the proton charge radius puzzle \cite{CREMA}.
There exists a number of attempts to reconsider a calculation of nuclear structure corrections in
\cite{structure} (see also other references in \cite{CREMA,borie1}) accounting among other things
the off-shell effects in two-photon exchange amplitudes.
In this work we study the corrections of special kind of order $\alpha(Z\alpha)^5$ related with the finite
size of the proton and deuteron in the hyperfine structure of muonic hydrogen. Preliminary estimate of
possible value of such contribution to hyperfine splitting for muonic hydrogen, as an example,
gives the numerical value $\alpha^2E_F(\mu p)\approx 0.011$ meV. This means that present corrections can be
important in order to obtain hyperfine splittings of $S$-states with high accuracy. For precise determination
of order $\alpha(Z\alpha)^5$ contribution we should take into account that the distributions of the charge and
magnetic moment of nuclei are described by electromagnetic form factors.

Our calculation is performed on the basis of quasipotential method in quantum electrodynamics (QED)
as applied to particle bound states, which was used previously for the solution of different
problems \cite{faustov}. In terms of perturbation theory in QED the contribution to the scattering
amplitude and quasipotential is determined by the Feynman diagrams presented in Fig.1.
To evaluate corrections of order $\alpha(Z\alpha)^5$ we neglect relative momenta of particles
in initial and final states and construct separate hyperfine potentials corresponding to muon self-energy,
vertex and spanning photon diagrams.

\begin{figure}[t!]
\centering
\includegraphics{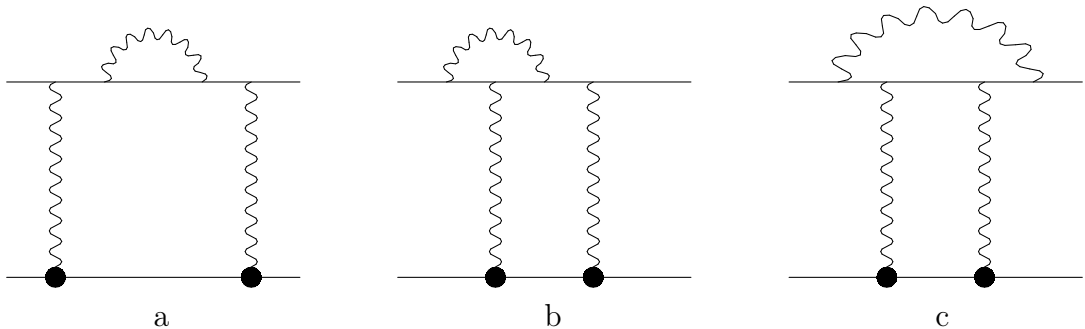}
\caption{Direct two-photon exchange amplitudes with radiative corrections to muon line
giving contributions of order $\alpha(Z\alpha)^5$ to the hyperfine structure. Wave line on the diagram denotes
the photon. Bold point on the diagram denotes the vertex operator of the proton or deuteron.}
\label{fig:pic1}
\end{figure}

Basic contribution to hyperfine splitting of $S$-states in muonic deuterium (below we present general equations
for muonic deuterium) is given by the following one-photon potential:
\begin{equation}
\label{eq:hfs}
\Delta V^{hfs}=\frac{4\pi\alpha}{3m_1m_p}\mu_d({\bf s}_1\cdot{\bf s}_2)\delta({\bf x}),
\end{equation}
where ${\bf s}_{1,2}$ are the spin operators of the muon and deuteron, $\mu_d=0.8574382308$ is magnetic moment
of the deuteron in nuclear magnetons, $m_1$ and $m_p$ are the masses of the muon and deuteron correspondingly.
Averaging \eqref{eq:hfs} over the bound state wave functions we obtain the leading order contribution to hyperfine
splitting (the Fermi energy) in muonic deuterium:
\begin{equation}
\label{eq:fermi}
\Delta E_F(S)=\frac{2\mu^3\alpha^4\mu_d}{m_1m_p n^3}=\Biggl\{{{49.0875~{\rm meV},~n=1}\atop{~~6.1359~{\rm meV},~n=2}}.
\end{equation}

The contribution of two-photon exchange diagrams to hyperfine structure of order $(Z\alpha)^5$ was investigated
earlier by many authors \cite{egs}. The lepton line radiative corrections to two-photon exchange amplitudes
were studied in detail in \cite{eides1} in the case of muonium hyperfine splitting. Total integral expression
for all radiative corrections in Fig.1 to hyperfine split ting of order  $\alpha(Z\alpha)^5$ including recoil
effects was obtained in \cite{eides2} in the Fried-Yennie gauge for radiative photon \cite{fried}.
The advantage of the Fried-Yennie gauge in the analysis of corrections in Fig.1 consists in the fact that
it leads to infrared-finite renormalizable integral expressions for muon self-energy operator, vertex function and
lepton tensor describing the "jellyfish-type" diagram (with spanning photon) \cite{eides3}. Using such expressions
we can perform analytical calculation of order $\alpha(Z\alpha)^5$ corrections to hyperfine structure
in the point-like nucleus approximation. If the approximation of point-like nucleus is inappropriate then these
expressions allow to obtain numerical values of diagrams (a,b,c) in Fig.1 separately. In this work we perform
independent construction of all enumerated above muon radiative corrections in the Fried-Yennie gauge and obtain
new integral contributions for the muon self-energy, vertex and spanning photon amplitudes separately to hyperfine structure
in the case of finite size nucleus. The muon-deuteron scattering amplitude can be presented in the form
(direct two-photon exchange diagrams with radiative corrections to the muon line):
\begin{equation}
\label{eq:2gamma}
{\cal M}=\frac{-i(Z\alpha)^2}{\pi^2}\int d^4k
\left[
\bar u(q_1)L_{\mu\nu}u(p_1)\right]D_{\mu\omega}(k)D_{\nu\lambda}(k)\times
\end{equation}
\begin{displaymath}
\left[\epsilon^\ast_\rho(q_2)\Gamma_{\omega_,\rho\beta}(q_2,p_2+k){\cal D}_{\beta\tau}(p_2+k)\Gamma_{\lambda,\tau\alpha}(p_2+k,p_2)
\epsilon_\alpha(p_2)\right]
\end{displaymath}
where $\epsilon^\ast_\rho(q_2)$ ($\epsilon_\alpha(p_2)$) denote the polarization vector of the final (initial) deuteron,
$p_{1,2}$ and $q_{1,2}$ are four-momenta of the muon and
deuteron in initial and final states: $p_{1,2}\approx q_{1,2}$. $k$ stands for the four-momentum of the exchange photon.
The vertex operator describing the photon-deuteron interaction
is determined by three electromagnetic form factors in the form
\begin{equation}
\label{eq:gamma}
\Gamma_{\omega_,\rho\beta}(q_2,p_2+k)=\frac{(2p_2+k)_\omega}{2m_2}g_{\rho\beta}\cdot F_1(k)-\frac{(2p_2+k)_\omega}{2m_2}
\frac{k_{\rho}k_{\beta}}{2m_2^2}\cdot F_2(k)-(g_{\rho\gamma}g_{\beta\omega}-
g_{\rho\omega}g_{\beta\gamma})\frac{k_\gamma}{2m_2}\cdot F_3(k).
\end{equation}
The form factors $F_{1,2,3}(k^2)$ are related to the
charge, magnetic and quadrupole deuteron form factors as ($\eta =k^2/4m_2^2$)
\begin{equation}
\label{eq:fcfmfq}
F_C=F_1+\frac{2}{3}\eta\left[f_1+(1+\eta)F_2-F_3\right],~~~F_M=F_3,~~~F_Q=F_1+(1+\eta)F_2-F_3.
\end{equation}
The propagators of the deuteron and photon in the Coulomb gauge are the following ones:
\begin{equation}
\label{eq:prop}
{\cal D}_{\alpha\beta}(p)=\frac{-g_{\alpha\beta}+\frac{p_\alpha p_\beta}{m_2^2}}{(p^2-m_2^2+i0)},~~~
D_{\lambda\sigma}(k)=\frac{1}{k^2}\left[g_{\lambda\sigma}+\frac{k_\lambda k_\sigma-k_0 k_\lambda g_{\sigma 0}-
k_0 k_\sigma g_{\lambda 0}}{{\bf k}^2}\right].
\end{equation}
The lepton tensor $L_{\mu\nu}$ has a completely definite form for each amplitude in Fig.1.
It is equal to the sum of the self-energy ($\Sigma$), vertex ($\Lambda$), and spanning photon ($\Xi$) insertions in the muon line:
\begin{equation}
\label{eq:lepton}
L_{\mu\nu}=L_{\mu\nu}^\Sigma+2L_{\mu\nu}^\Lambda+L_{\mu\nu}^\Xi.
\end{equation}
Using the FeynCalc package \cite{fc} we construct the renormalized muon self-energy operator and vertex
function as in \cite{eides2,eides3} and obtain
the following expressions for leptonic tensors corresponding to
muon self-energy, vertex contributions and the diagram with spanning photon in the Fried-Yennie gauge:
\begin{equation}
\label{eq:se}
L_{\mu\nu}^{\Sigma}=-\frac{3\alpha}{4\pi}\gamma_\mu(\hat p_1-\hat k)\gamma_\nu\int_0^1\frac{(1-x)dx}{(1-x)m_1^2+x{\bf k}^2},
\end{equation}
\begin{equation}
\label{eq:vertex}
L_{\mu\nu}^{\Lambda}=\frac{\alpha}{4\pi}\int_0^1 dz\int_0^1 dx \gamma_\mu
\frac{\hat p_1-\hat k+m_1}{(p_1-k)^2-m_1^2+i0}
\left[F_\nu^{(1)}+\frac{F_\nu^{(2)}}{\Delta}+\frac{F_\nu^{(3)}}{\Delta^2}\right],
\end{equation}
\begin{equation}
\label{eq:verfun}
F_\nu^{(1)}=-6x\gamma_\nu\ln\frac{m_1^2x+{\bf k}^2z(1-xz)}{m_1^2x},~~~F_\nu^{(3)}=2x^3(1-x)\hat Q(\hat p_1-
\hat k+m_1)\gamma_\nu(\hat p_1+m_1)\hat Q,
\end{equation}
\bga
F_\nu^{(2)}=-x^3(2\gamma_\nu Q^2-2\hat Q\gamma_\nu \hat Q)-x^2[\gamma_\alpha\hat Q\gamma_\nu(\hat p_1+m_1)\gamma_\alpha+
\gamma_\alpha(\hat p_1-\hat k+m_1)\gamma_\nu\hat Q\gamma_\alpha+\\
2\gamma_\nu(\hat p_1+m_1)\hat Q
+2\hat Q(\hat p_1-\hat k+m_1)\gamma_\nu]-x(2-x)\gamma_\alpha(\hat p_1-\hat k+m_1)\gamma_\nu(\hat p_1+m_1)\gamma_\alpha,
\ega
\begin{displaymath}
Q=-p_1+kz,~\Delta=x^2m_1^2-xz(1-xz)k^2+2kp_1xz(1-x),
\end{displaymath}
\begin{equation}
\label{eq:jellyfish}
L_{\mu\nu}^{\Xi}=\frac{\alpha}{4\pi}\int_0^1 dz\int_0^1 dx \left(\frac{F^{(1)}_{\mu\nu}}{\Delta}+
\frac{F^{(2)}_{\mu\nu}}{\Delta^2}+\frac{F^{(3)}_{\mu\nu}}{\Delta^3}\right),
\end{equation}
\begin{equation}
F^{(1)}_{\mu\nu} =12 \hat p_1 g_{\mu\nu}-15\hat p_1\gamma_\mu\gamma_\nu
+9\hat k\gamma_\mu\gamma_\nu
-24x\hat Q\gamma_\mu\gamma_\nu+16x\hat Qg_{\mu\nu}-8x\gamma_\mu Q_\nu+\gamma_\nu(30p_{1,\mu}-18k_\mu +40Q_\mu x),
\end{equation}
\begin{equation}
F^{(2)}_{\mu\nu} =\hat p_1\hat Q\gamma_\mu (  - 2p_{1,\nu}x - 8Q_\nu x^2 )
+ \hat p_1\hat Q\gamma_\nu ( 2p_{1,\mu}x+8Q_\mu x^2)+\hat p_1\gamma_\mu\gamma_\nu(- 8m_1^2x + 5Q^2x^2)\\
\end{equation}
\begin{displaymath}
+\hat p_1\gamma_\mu  ( 2Q_\nu m_1x )+\hat p_1\gamma_\nu(-2Q_\mu m_1x )+\hat p_1(12p_{1,\mu}p_{1,\nu}-
12p_{1,\mu}k_\nu+20p_{1,\mu}Q_\nu x+ 24p_{1,\nu}Q_\mu x - 12k_\mu Q_\nu x\\
\end{displaymath}
\begin{displaymath}
-12k_\nu Q_\mu x+32Q_\mu Q_\nu x^2 + 16g_{\mu\nu}m_1^2x + 12g_{\mu\nu}kQx-12g_{\mu\nu}Q^2x^2)
+ 6\hat k\hat p_1\gamma_\mu p_{1,\nu} -6\hat k\hat p_1\gamma_\nu p_{1,\mu}\\
\end{displaymath}
\begin{displaymath}
+ \hat k\hat Q\gamma_\mu  ( 12p_{1,\nu}x + 8Q_\nu x^2)
+ \hat k\hat Q\gamma_\nu  (  - 12p_{1,\mu}x - 8Q_\mu x^2)
-3\hat k\gamma_\mu\gamma_\nu  Q^2x^2 -6 \hat k\gamma_\mu  Q_\nu m_1x\\
\end{displaymath}
\begin{displaymath}
+ \hat k\gamma_\nu  ( 6Q_\mu m_1x )+ \hat k  ( 12p_{1,\mu}Q_\nu x - 12p_{1,\nu}Q_\mu x + 12g_{\mu\nu}m_1^2x -
8g_{\mu\nu}Q^2x^2 )+ \hat Q\gamma_\mu\gamma_\nu  ( 2m_1^2x - 24m_1^2x^2 -\\
\end{displaymath}
\begin{displaymath}
8kQx^2+16Q^2x^3)+ \hat Q\gamma_\mu  (2p_{1,\nu}m_1x+8Q_\nu m_1x^2)+\hat Q\gamma_\nu(- 2p_{1,\mu}m_1x-8Q_\mu m_1x^2 )
+ \hat Q  (-4p_{1,\mu}p_{1,\nu}x-\\
\end{displaymath}
\begin{displaymath}
12p_{1,\mu}k_\nu x + 12p_{1,\nu}
k_\mu x + 16p_{1,\nu}Q_\mu x^2 - 16k_\nu Q_\mu x^2 + 20Q_\mu Q_\nu x^3 +
16g_{\mu\nu}m_1^2x^2 + 16g_{\mu\nu}kQx^2 - 14g_{\mu\nu}Q^2 x^3 )\\
\end{displaymath}
\begin{displaymath}
+g_{\mu\nu}(-2m_1^3x-6kQm_1x + 4Q^2m_1x^2)+\gamma_\mu(4p_{1,\nu}m_1^2x -12p_{1,\nu}kQx -8p_{1,\nu}Q^2x^2 -
12k_\nu m_1^2x+8k_\nu Q^2x^2 +\\
\end{displaymath}
\begin{displaymath}
2Q_\nu m_1^2x - 16Q_\nu m_1^2x^2-16Q_\nu kQx^2 + 2Q_\nu Q^2x^3)+\gamma_\nu(24p_{1,\mu}m_1^2x +
12p_{1,\mu}kQx - 18p_{1,\mu}Q^2x^2\\
\end{displaymath}
\begin{displaymath}
-12k_\mu m_1^2x + 14k_\mu Q^2x^2 - 2Q_\mu m_1^2x + 48Q_\mu m_1^2x^2+16Q_\mu kQx^2-30Q_\mu Q^2x^3)
- 24p_{1,\mu}p_{1,\nu}m_1+\\
\end{displaymath}
\begin{displaymath}
12p_{1,\mu}k_\nu m_1-36p_{1,\mu}Q_\nu m_1x + 12
p_{1,\nu}k_\mu m_1 - 36p_{1,\nu}Q_\mu m_1x + 24k_\mu Q_\nu m_1x + 12k_\nu Q_\mu m_1x-\\
\end{displaymath}
\begin{displaymath}
48Q_\mu Q_\nu m_1x^2- 16g_{\mu\nu}m_1^3x - 12g_{\mu\nu}kQm_1x + 8g_{\mu\nu}Q^2m_1x^2,
\end{displaymath}
\begin{equation}
F^{(3)}_{\mu\nu}=\hat p_1\hat Q\gamma_\mu  ( 8p_{1,\nu}kQx^2 - 8Q_\nu m_1^2x^3 + 4Q_\nu Q^2x^4 )
+ \hat p_1\hat Q\gamma_\nu  (  - 8p_{1,\mu}kQx^2 + 8Q_\mu m_1^2x^3 - 4Q_\mu
Q^2x^4 )+
\end{equation}
\begin{displaymath}
\hat p_1(- 16p_{1,\mu}p_{1,\nu}Q^2x^2 + 8p_{1,\mu}k_\nu Q^2x^2 +
16p_{1,\mu}Q_\nu kQx^2 - 16p_{1,\mu}Q_\nu Q^2x^3 + 8p_{1,\nu}k_\mu
Q^2x^2 - 16p_{1,\nu}Q_\mu kQx^2 -
\end{displaymath}
\begin{displaymath}
16p_{1,\nu}Q_\mu Q^2x^3 + 8k_\mu Q_\nu Q^2x^3 + 8k_\nu Q_\mu Q^2x^3 - 16Q_\mu Q_\nu Q^2x^4 - 8
g_{\mu\nu}kQQ^2x^3 - 8g_{\mu\nu}Q^2m_1^2x^3 + 4g_{\mu\nu}Q^4x^4 )
\end{displaymath}
\begin{displaymath}
+\hat k\hat p_1\hat Q  ( 8p_{1,\mu}Q_\nu x^2 - 8p_{1,\nu}Q_\mu x^2 )
+ \hat k\hat Q\gamma_\mu  ( 8p_{1,\nu}m_1^2x^2 - 8p_{1,\nu}Q^2x^3 - 4Q_\nu Q^2x^4 )
+ \hat k\hat Q\gamma_\nu  (  - 8p_{1,\mu}m_1^2x^2 +
\end{displaymath}
\begin{displaymath}
8p_{1,\mu}Q^2x^3 + 4Q_\mu Q^2x^4 )+ \hat k\hat Q  ( 8p_{1,\mu}Q_\nu m_1x^2 - 8p_{1,\nu}Q_\mu m_1x^2 )
+ \hat k\gamma_\mu  ( 4p_{1,\nu}Q^2m_1x^2 + 4Q_\nu Q^2m_1x^3 )
\end{displaymath}
\begin{displaymath}
+ \hat k\gamma_\nu  (  - 4p_{1,\mu}Q^2m_1x^2 - 4Q_\mu Q^2m_1x^3 )
+ \hat k  ( 16p_{1,\mu}Q_\nu m_1^2x^2 - 8p_{1,\mu}Q_\nu Q^2x^3 - 16
p_{1,\nu}Q_\mu m_1^2x^2 +
\end{displaymath}
\begin{displaymath}
8p_{1,\nu}Q_\mu Q^2x^3 - 8g_{\mu\nu}Q^2m_1^2
x^3 + 4g_{\mu\nu}Q^4x^4 )+ \hat Q\gamma_\mu\gamma_\nu  (  - 8m_1^4x^3 + 4kQQ^2x^4 + 12Q^2m_1^2x^4 - 4
Q^4x^5 ) +
\end{displaymath}
\begin{displaymath}
\hat Q\gamma_\mu  (  - 8p_{1,\nu}kQm_1x^2 + 8Q_\nu m_1^3x^3 - 4Q_\nu Q^2m_1x^4 )
+ \hat Q\gamma_\nu  ( 8p_{1,\mu}kQm_1x^2 - 8Q_\mu m_1^3x^3 + 4Q_\mu Q^2m_1x^4 )+
\end{displaymath}
\begin{displaymath}
\hat Q  (  - 16p_{1,\mu}p_{1,\nu}m_1^2x^2 + 8p_{1,\mu}k_\nu Q^2x^3 -
16p_{1,\mu}Q_\nu m_1^2x^3 + 16p_{1,\nu}k_\mu m_1^2x^2 - 8p_{1,\nu}k_\mu
Q^2x^3 - 8p_{1,\nu}Q_\mu Q^2x^4 +
\end{displaymath}
\begin{displaymath}
8k_\nu Q_\mu Q^2x^4 - 8Q_\mu
Q_\nu Q^2x^5 - 8g_{\mu\nu}kQQ^2x^4 - 8g_{\mu\nu}Q^2m_1^2x^4 +
4g_{\mu\nu}Q^4x^5 )+ g_{\mu\nu}  ( 4kQQ^2m_1x^3 + 4Q^2m_1^3x^3
\end{displaymath}
\begin{displaymath}
-2Q^4m_1x^4 )+\gamma_\mu  ( 8p_{1,\nu}kQQ^2x^3 - 8p_{1,\nu}Q^2m_1^2x^3 + 4p_{1,\nu}
Q^4x^4 + 8k_\nu Q^2m_1^2x^3 - 4k_\nu Q^4x^4 - 16Q_\nu m_1^4x^3 +
\end{displaymath}
\begin{displaymath}
8Q_\nu kQ Q^2x^4 + 8Q_\nu Q^2m_1^2x^4 )+\gamma_\nu (-8p_{1,\mu}kQQ^2x^3-8p_{1,\mu}Q^2m_1^2x^3 +4
p_{1,\mu}Q^4x^4+8k_\mu Q^2m_1^2x^3 -
\end{displaymath}
\begin{displaymath}
4k_\mu Q^4x^4 + 16Q_\mu m_1^4x^3 - 8Q_\mu kQ Q^2x^4-24Q_\mu Q^2m_1^2x^4 + 8Q_\mu Q^4x^5)
+24p_{1,\mu}p_{1,\nu}Q^2m_1x^2 -
\end{displaymath}
\begin{displaymath}
8p_{1,\mu}k_\nu Q^2m_1x^2 - 16p_{1,\mu}
Q_\nu kQm_1x^2 + 24p_{1,\mu}Q_\nu Q^2m_1x^3 - 16p_{1,\nu}k_\mu Q^2
m_1x^2 + 16p_{1,\nu}Q_\mu kQm_1x^2 +
\end{displaymath}
\begin{displaymath}
24p_{1,\nu}Q_\mu Q^2m_1x^3 - 16
k_\mu Q_\nu Q^2m_1x^3 - 8k_\nu Q_\mu Q^2m_1x^3 + 24Q_\mu Q_\nu
Q^2m_1x^4 + 8g_{\mu\nu}kQQ^2m_1x^3 +
\end{displaymath}
\begin{displaymath}
8g_{\mu\nu}Q^2m_1^3x^3 - 4g_{\mu\nu}Q^4m_1x^4.
\end{displaymath}
For the further construction of hyperfine splitting potentials corresponding to the amplitude \eqref{eq:2gamma}
we introduce the projection operators on the states of muon-deuteron pair with the spin 3/2 and 1/2:
\begin{equation}
\label{eq:project}
\hat\Pi_{\mu,3/2}=[u(p_1)\epsilon_\mu(p_2)]_{3/2}=\Psi_\mu(P),~~~\hat\Pi_{\mu,1/2}=\frac{i}{\sqrt{3}}\gamma_5\left(\gamma_\mu-
v_{1,\mu}\right)\Psi(P),
\end{equation}
\begin{equation}
\label{eq:sum}
\sum_{\lambda}\Psi_\mu^\lambda(P)\bar\Psi_\nu^\lambda(P)=-\frac{\hat v_1+1}{2}
\left(g_{\mu\nu}-\frac{1}{3}\gamma_\mu\gamma_\nu-\frac{2}{3}v_{1,\mu}v_{1,\nu}+\frac{1}{3}
(v_{1,\mu}\gamma_\nu-v_{1,\nu}\gamma_\mu\right),
\end{equation}
where the spin-vector $\Psi_\mu(P)$ and spinor $\Psi(P)$ describe the muon-deuteron bound states with
spins 3/2 and 1/2, $v_{1,\mu}=P_\mu/M$, $P=p_1+p_2$, $M=m_1+m_2$. The insertion \eqref{eq:project} into \eqref{eq:2gamma}
allows us to pass to the trace calculation and contractions over the Lorentz indices by means of the system
Form \cite{form}. A general structure of potentials contributing to the energy shifts for states with the angular momenta
1/2 and 3/2 is the following one:
\begin{equation}
\label{eq:pot12}
N_{1/2}=\frac{1}{6}Tr\Bigl\{\sum_{\sigma}\Psi^\sigma(P)\bar\Psi^\sigma(P)(\gamma_\rho-v_{1,\rho})\gamma_5(1+\hat v_1)
L_{\mu\nu}(1+\hat v_1)\gamma_5(\gamma_\alpha-v_{1,\alpha})\Bigr\}\times
\end{equation}
\begin{displaymath}
\Gamma_{\omega,\rho\beta}(q_2,p_2+k){\cal D}_{\beta\tau}(p_2+k)\Gamma_{\lambda,\tau\alpha}(p_2+k,p_2)D_{\mu\omega}(k)D_{\nu\lambda}(k),
\end{displaymath}
\begin{equation}
\label{eq:pot32}
N_{3/2}=\frac{1}{4}Tr\Bigl\{\sum_{\sigma}\Psi^\sigma_\alpha(P)\bar\Psi^\sigma_\rho(P)(1+\hat v_1)
L_{\mu\nu}(1+\hat v_1)\Bigr\}\times
\end{equation}
\begin{displaymath}
\Gamma_{\omega,\rho\beta}(q_2,p_2+k){\cal D}_{\beta\tau}(p_2+k)\Gamma_{\lambda,\tau\alpha}(p_2+k,p_2)
D_{\mu\omega}(k)D_{\nu\lambda}(k).
\end{displaymath}
The expressions \eqref{eq:pot12} and \eqref{eq:pot32} contain both recoil and nonrecoil corrections
of order $\alpha(Z\alpha)^5$.
Since we neglect the recoil effects the denominator of the deuteron propagator is simplified as follows:
$1/[(p_2+k)^2-m_2^2+i0]\approx 1/(k^2+2kp_2+i0)\approx 1/(2k_0m_2+i0)$. The crossed two-photon amplitudes give in this case
a similar contribution to hyperfine splitting which is determined also by relations \eqref{eq:2gamma}-\eqref{eq:jellyfish}
with the replacement $k\to -k$ in the deuteron propagator. As a result the summary contribution is proportional to the
$\delta(k_0)$:
\begin{equation}
\label{eq:delta}
\frac{1}{2m_2k_0+i0}+\frac{1}{-2m_2k_0+i0}=-\frac{i\pi}{m_2}\delta(k_0).
\end{equation}
In the case of muonic hydrogen the transformation of the scattering amplitude and a construction
of muon-proton potential can be done in much the same way. The main difference is related with
the structure of proton-photon vertex functions which are parameterized by two electromagnetic form
factors. Another difference appears in the projection operators on the states with spin 1 and 0
which have the form:
\begin{equation}
\label{eq:proj_mup}
\hat\Pi_{0,1}=\frac{\hat v_1+1}{2\sqrt{2}}\gamma_5(\hat\epsilon),
\end{equation}
where $\epsilon_\mu$ is the polarization vector of muon-proton state with spin 1.
The energy shift caused by interactions shown in Fig.1 is given by
\begin{equation}
\label{eq:shift}
\Delta E_{1/2,3/2}={\cal M}_{1/2,3/2}|\psi_n(0)|^2,
\end{equation}
where $|\psi_n(0)|^2=(\mu Z\alpha)^3/\pi n^3$ is the squared modulus of the bound state wave function
at the origin. The lower subscript denotes total angular momentum for the muon-deuteron state.
Then the hyperfine splitting (hfs) is determined as follows:
\begin{equation}
\label{eq:dehfs}
\Delta E^{hfs}=\Delta E_{3/2}-\Delta E_{1/2}.
\end{equation}

As a result three types of corrections of order  $\alpha(Z\alpha)^5$ to hyperfine structure
in both cases of muonic hydrogen are presented in the integral form over the loop momentum ${\bf k}$
and the Feynman parameters $x$ and $z$:
\begin{equation}
\label{eq:result1}
\Delta E^{hfs}_{\Sigma}=E_F6\frac{\alpha(Z\alpha)}{\pi^2}\int_0^1 xdx\int_0^\infty\frac{F_1(k^2)F_3(k^2)dk}{x+(1-x)k^2},
\end{equation}
\begin{equation}
\label{eq:result21}
\Delta E^{hfs}_{\Lambda~1}=-E_F24\frac{\alpha(Z\alpha)}{\pi^2}\int_0^1 dz\int_0^1 xdx\int_0^\infty
\frac{F_1(k^2)F_3(k^2)\ln [\frac{x+k^2z(1-xz)}{x}]dk}{k^2},
\end{equation}
\begin{equation}
\label{eq:result22}
\Delta E^{hfs}_{\Lambda~2}=E_F8\frac{\alpha(Z\alpha)}{\pi^2}\int_0^1 dz\int_0^1 dx\int_0^\infty\frac{dk}{k^2}
\Biggl\{\frac{F_1(k^2)F_3(k^2)}{[x+k^2z(1-xz)]^2}\Bigl[-2xz^2(1-xz)k^4+
\end{equation}
\begin{displaymath}
zk^2(3x^3z-x^2(9z+1)+x(4z+7)-4)+x^2(5-x)\Bigr]-\frac{1}{2}\Biggr\},
\end{displaymath}
\begin{equation}
\label{eq:result3}
\Delta E^{hfs}_{\Xi}=E_F4\frac{\alpha(Z\alpha)}{\pi^2}\int_0^1(1-z)dz\int_0^1(1-x)dx\int_0^\infty
\frac{F_1(k^2)F_3(k^2)dk}{[x+(1-x)k^2]^3}
\end{equation}
\begin{displaymath}
\times\Bigl[6x+6x^2-6x^2z+2x^3-12x^3z-12x^4z+k^2(-6z+18xz+4xz^2+7x^2z-30x^2z^2-
\end{displaymath}
\begin{displaymath}
2x^2z^3-36x^3z^2+12x^3z^3+24x^4z^3)+k^4(9xz^2-31x^2z^3+34x^3z^4-12x^4z^5\Bigr],
\end{displaymath}
where we extracted the value of the deuteron magnetic moment from $F_3(k^2)$ so that
$F_3(0)=1$ and $F_1(0)=1$. The dimensionless variable $k$ is introduced in \eqref{eq:result1}-\eqref{eq:result3}.
The contribution of the form factor $F_2(k^2)$ to \eqref{eq:result1}-\eqref{eq:result3}
is omitted because the terms $F_2(k^2)F_3(k^2)$ are suppressed by powers of the mass $m_2$.
The term $1/2$ in figure brackets \eqref{eq:result22} is related with the subtraction term of
the quasipotential. All corrections \eqref{eq:result1}, \eqref{eq:result21},
\eqref{eq:result22} and \eqref{eq:result3} are expressed through the convergent integrals.
In the case of point-like deuteron (proton) all integrations can be done analytically.
Firstly, the integration over the parameter $x$ is performed and after that the integration
over $k$ and $z$. The diagrams of the seagull type for point-like deuteron doesn't contribute
to hyperfine splitting.
In Table I we present separate results for muon self-energy, vertex and
spanning photon contributions in the Fried-Yennie gauge. Total analytical result equal to
$E_F\alpha(Z\alpha)(\ln 2-\frac{13}{4})$ was obtained for the first time in
\cite{kroll}. In \cite{eides3} the expressions for the lepton tensors
of the vertex and spanning photon diagrams were constructed in a slightly different form
but they lead to the same contributions \eqref{eq:result1}-\eqref{eq:result3}
to hyperfine splitting of $S$-states in the case of point-like nucleus.
In numerical calculations \eqref{eq:result1}-\eqref{eq:result3} with finite size nucleus
we employ the known parameterizations \cite{abbott,kelly} for electromagnetic form factors
of the deuteron and proton used also in our previous papers \cite{faustov1}.

\begin{table}[h]
\caption{\label{t1} Radiative nuclear finite size corrections of order $\alpha(Z\alpha)^5$,
to hyperfine structure of $S$-states in muonic hydrogen. Numerical results for the ground
state are presented. The contribution to the hyperfine structure for the point nucleus is
indicated in round brackets.}
\bigskip
\begin{tabular}{|c|c|c|c|c|}   \hline
Bound state  &  SE correction,  &   Vertex    & Spanning photon  &  Summary \\
           &        meV       &    correction, meV                    & contribution, meV & contribution, meV   \\    \hline
Point-like nucleus  &$E_F\alpha(Z\alpha)\frac{3}{2}$    & $-E_F\alpha(Z\alpha)(3\ln 2+\frac{9}{4})$     &   $E_F\alpha(Z\alpha)(4\ln 2-\frac{5}{2})$& $E_F\alpha(Z\alpha)(\ln 2-\frac{13}{4})$    \\   \hline
Muonic   &  0.0083  &   -0.0915   &  -0.0028  & -0.0860  \\
hydrogen     & (0.0146)   &  (-0.0421)    & (0.0026)   & (-0.0249)   \\   \hline
Muonic   &  0.0014  & -0.0042     & -0.0011   & -0.0039   \\
deuterium  & (0.0039)   &(-0.0113)      & (0.0007)  & (-0.0067)    \\   \hline
\end{tabular}
\end{table}

It follows from obtained results in Table~I that the account of proton and deuteron
form factors essentially changes the results for point-like nuclei. In a number
of cases there is the change of the correction sign. This follows from the fact that
for muonic atoms the integral over $k$ in \eqref{eq:result1}-\eqref{eq:result3} is specified
by the interval of order of muon mass and a sign-alternating integrand.
We perform
independent calculation of nonrecoil corrections of order $\alpha(Z\alpha)^5$ to hyperfine
structure of $S$-states in muonic hydrogen using the Fried-Yennie gauge for radiative photon.
In the case of muonic hydrogen these corrections decrease the theoretical value of hyperfine splitting
of $2S$-state approximately on 0.01 meV. To construct the quasipotential corresponding to
amplitudes in Fig.1 we develop the method of projection operators on the bound states with
definite spins. It allows to employ different systems of analytical calculations
\cite{form,fc}. In this approach more complicated corrections, for example, radiative recoil corrections
to hyperfine structure of order $\alpha(Z\alpha)^5m_1/m_2$ can be evaluated if an increase of the accuracy
will be needed. The results from Table~I should be taken into account to obtain total value
of hyperfine splittings in muonic hydrogen for a comparison with experimental data \cite{CREMA}.

We are grateful to F.~Kottmann for valuable information about CREMA experiments.
The work is supported by the Russian Foundation for Basic Research (grant 14-02-00173)
and Dynasty foundation.

\end{document}